\documentclass[useAMS,usenatbib]{mn2e}

\usepackage{graphicx}
\usepackage{txfonts}
\usepackage[breaklinks=true]{hyperref}
\usepackage{longtable}
\usepackage{subfig}
\usepackage{float}
\usepackage{booktabs}
\usepackage{natbib}
\usepackage{ amssymb }
\newcommand{\gsim}{\lower.7ex\hbox{$\;\stackrel{\textstyle>}{\sim}\;$}}

\def\araa{ARA\&A}%
\def\apj{ApJ}%
\def\apjl{ApJ}%
\def\aap{A\&A}%
\def\aapr{A\&A~Rev.}%
\def\mnras{MNRAS}%
\def\nar{New A Rev.}%
\def\nat{Nature}%

\title[X-ray time lags and nonlinear variability in ULXs]{X-ray time lags and nonlinear variability in the ultraluminous X-ray sources NGC\,5408 X-1 and NGC\,6946 X-1}
\author[Hern\'{a}ndez-Garc\'{i}a et al.]{L. Hern\'{a}ndez-Garc\'{i}a$^{1}$\thanks{E-mail:
lorena@iaa.es}, S. Vaughan$^{2}$, T.P. Roberts$^{3}$ and M. Middleton$^{4}$\\
$^{1}$Instituto de Astrof\'{i}sica de Andaluc\'{i}a, CSIC, Glorieta de
la Astronom\'{i}a, s/n, 18008 Granada, Spain\\
$^{2}$X-ray \& Observational Astronomy Group, Department of Physics and Astronomy, University of Leicester, Leicester LE1 7RH, UK\\
$^{3}$Centre for Extragalactic Astronomy, Department of Physics, Durham University, South Road, Durham DH1 3LE, UK\\
$^{4}$Institute of Astronomy, Madingley Rd, Cambridge, CB3 0HA, UK}
\begin{document}

\date{Draft:\today}

\pagerange{\pageref{firstpage}--\pageref{lastpage}} \pubyear{2015}

\maketitle

\label{firstpage}

\begin{abstract}
We present our analysis of the X-ray variability of two ultraluminous X-ray sources (ULXs) based on multiple {\it XMM--Newton} observations. We show the linear rms-flux relation is present in eight observations of NGC\,5408 X-1 and also in three observations of NGC\,6946 X-1, but data from other ULXs are generally not sufficient to constrain any rms-flux relation. The presence of this relation was previously reported in only two observations of NGC 5408\,X-1; our results show this is a persistent property of the variability of NGC\,5408 X-1 and extends to at least one other variable ULX. We speculate this is a ubiquitous property of ULX variability, as it is for X-ray variability in other luminous accreting sources. We also recover the time delay between hard and soft bands in NGC\,5408 X-1, with the soft band ($<$1 keV) delayed with respect to the hard band ($>$1 keV) by up to $\sim$10 s ($\sim$0.2 rad) at frequencies above $\sim$few mHz. For the first time, we extend the lag analysis to lower frequencies and find some evidence for a reversal of the lag, a hard lag of $\sim$1 ks at frequencies of $\sim$0.1 mHz. Our energy-resolved analysis shows the time delays are energy dependent.
We argue that the lag is unlikely to be a result of reflection from an accretion disc (`reverberation') based on the lack of reflection features in the spectra, and the large size of the reflector inferred from the magnitude of the lag. We also argue that associating the soft lag with a quasi-periodic oscillation (QPO) in these ULXs -- and drawing an analogy between soft lags in ULXs and soft lags seen in some low-frequency QPOs of Galactic X-ray binaries -- is premature. 
\end{abstract}


\section{Introduction}

In recent years substantial progress has been made in understanding the nature of ultraluminous X-ray sources (ULXs).  These objects are X-ray sources located within, but displaced from the nucleus of, nearby galaxies, that display observed X-ray luminosities in excess of $10^{39} ~\rm ~erg~s^{-1}$ \citep[see][for the most recent review]{feng2011}. In particular, there is strong evidence for three relatively nearby ULXs containing stellar-mass black holes ($M_{\rm BH} \sim 10 M_{\odot}$), and so accreting at super-Eddington rates \citep{middleton2013,liu2013, motch2014}. The last of these objects is particularly important as it directly links super-Eddington emission to the peculiar X-ray spectrum displayed by many bright ULXs \citep[see e.g.,][]{stobbart2006,bachetti2013}, and so supports the notion of some ULXs accreting in a new, super-Eddington {\it ultraluminous state\/} \citep{gladstone2009, sutton2013}. The physics of this state are only just emerging, but it appears that the variety of X-ray spectra and the coarse variability properties of individual objects may be consistent with a model in which massive, radiatively-driven winds are launched from a geometrically thick accretion disc, bloated by the advection of its hot inner regions towards the black hole, and a combination of the collimation of the X-radiation emerging from the innermost regions of the accretion flow by the optically thick wind and the viewing angle of the observer dictates what is seen \citep{poutanen2007, king2009, sutton2013, middleton2015}. However, this may not explain all ULXs, which we now are certain to be a heterogeneous population after the discovery of pulsations from a luminous ULX in M\,82, demonstrating that it hosts a neutron star \citep{bachetti2014}.  Additionally, there is evidence that at least some of the brightest ULXs may still harbour the long-sought intermediate-mass black holes \citep[$M_{\rm BH} \sim 100 - 10000 M_{\odot}$, e.g.][]{farrell2009, sutton2012, mezcua2015}. Much work therefore remains to be done to understand both the composition and the accretion physics of the ULX population.

\begin{figure*}
\includegraphics[width=0.35\textwidth,angle=90]{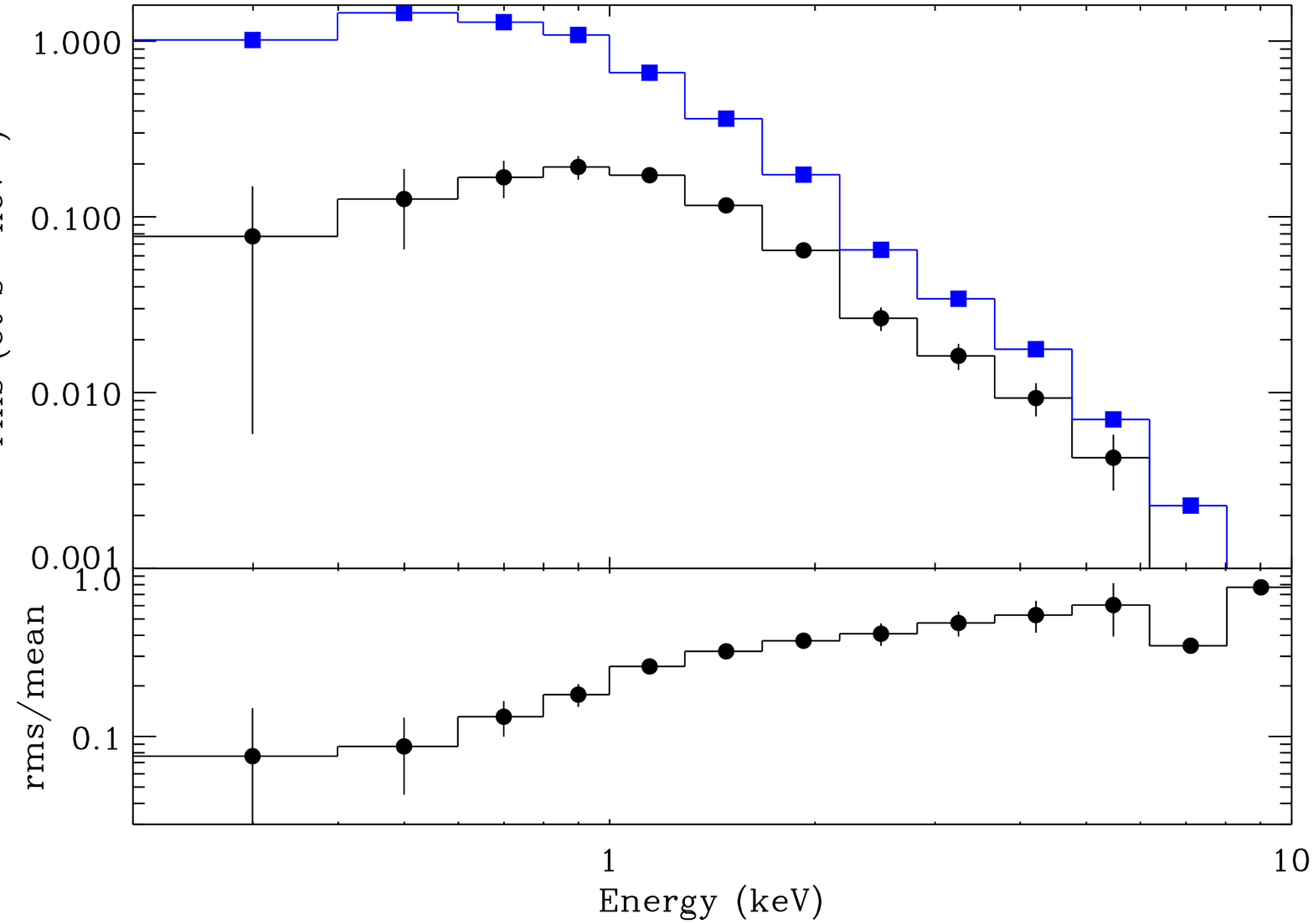}
\includegraphics[width=0.35\textwidth,angle=90]{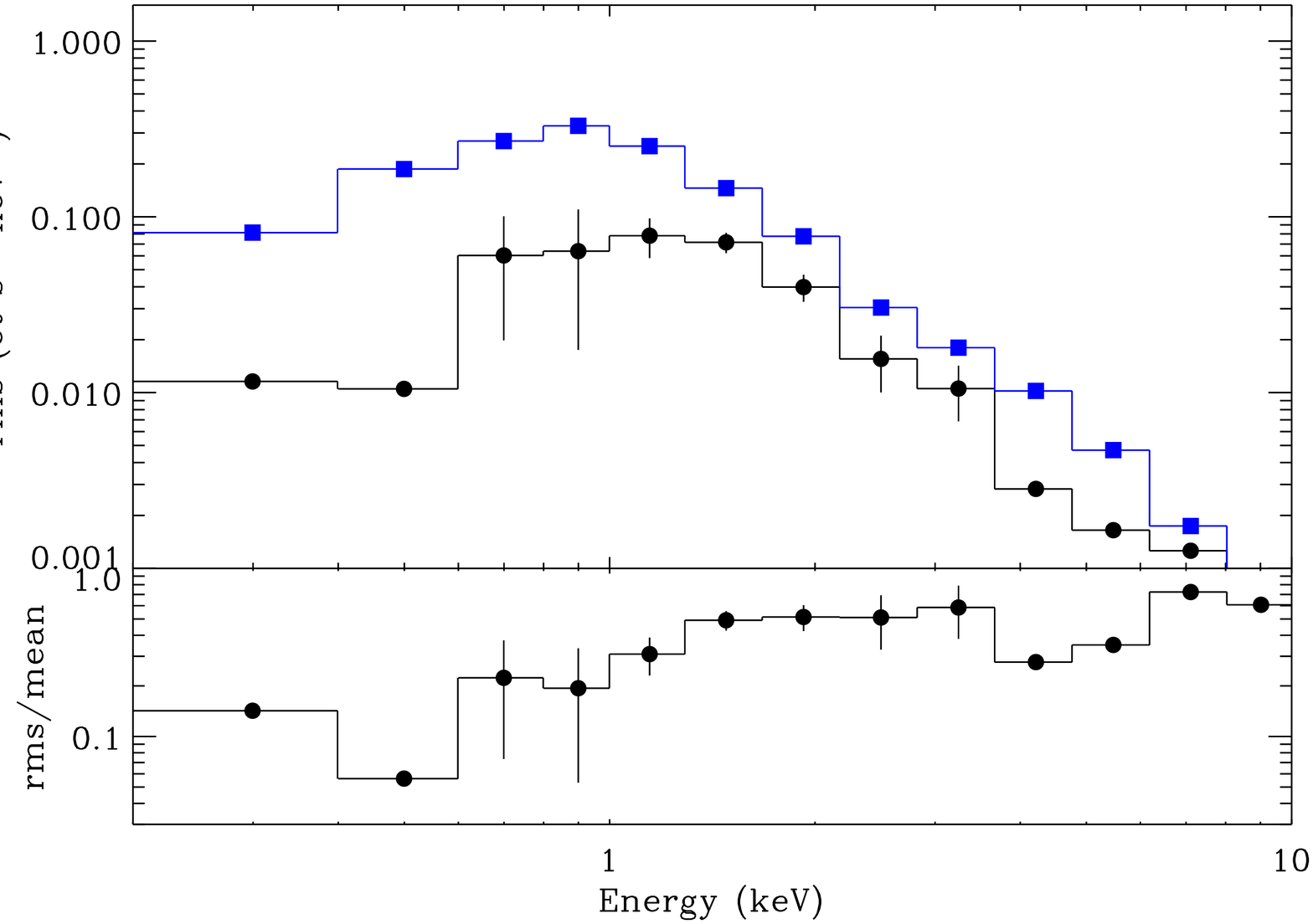}
\caption{rms spectrum (black circles) and mean spectrum (blue squares) of (left): NGC\,5408 X-1 using the 2010 and 2011 observations, and (right): NGC\,6946 X-1 using the 2007 and 2012 observations. The lower panels show the ratio between the rms and the mean spectrum, i.e., the fractional rms spectrum. We used a time bin size of 10.4 s respecting the ``good'' time interval list, and segments of 10 ksec duration for the estimation of the rms (see text). }
\label{rms}
\end{figure*}

X-ray variability can be a powerful tool for investigating luminous, accreting black-hole binaries \citep[BHB,][]{remillard2006} or active galactic nuclei \citep[AGN,][]{vaughan2003}. 
Studies of X-ray variability have been undertaken for a few ULXs using different approaches, from changes in their light curves \citep{sutton2013}, the study of quasi-periodic oscillations \citep[QPO, e.g.,][]{pasham2012,caballerogarcia2013}, the rms-flux relation or time delays between different energy bands \citep{heilvaughan2009, demarco2013b}. 
However, the study of variability in ULXs is hampered by their relatively low count rates, and usually requires long exposures in order to recover the statistical properties of their variability. For example, the rms-flux relation or time lags have to-date been detected in only one ULX, NGC\,5408 X-1. \citet{heilvaughan2010} first showed that this source follows the rms-flux relation from two {\it XMM--Newton} observations. This property of the X-ray variability seems to be ubiquitous among Galactic black hole X-ray binaries (XRB) and also AGN \citep{uttley2001, uttley2005, heil2012}. The discovery in NGC\,5408 X-1 demonstrates a strong connection between at least this ULX and the better-understood accreting black hole systems in XRBs and AGN. \citet{heilvaughan2010} also studied time delays between the soft and hard X-ray energy bands in NGC\,5408 X-1 and found a soft lag, i.e., variations in the soft photons lag those in hard photons at mHz frequencies. This result was later confirmed by \citet{demarco2013b}, who used six {\it XMM--Newton} observations to study the QPO and the soft time lag.

Here we revisit the {\it XMM--Newton} observations of ULXs, concentrating on NGC\,5408 X-1 and NGC\,6946 X-1. Our focus is on the frequency dependent time lags (extending to lower frequencies than previously published) and confirming and extending the one published example of the rms-flux relation in a ULX. 
This paper is organised as follows: in Sect. \ref{sample} we describe the data used for the analysis and the data reduction, in Sect. \ref{rmsflux} we present the rms-flux relation for the two ULXs, and in Sect. \ref{cross} the cross-spectral analysis and the results of the coherence, time and phase delay, and the phase lag spectrum. Finally, the results derived from this study are discussed in Sect. \ref{discussion}.

\begin{table}
 \caption{\label{obs}Observational details.}
 \begin{tabular}{@{}lcccc}
  \hline
Object & ObsID & Date & T$^a$ & T$^b$ \\
       &       &      & (ksec) & (ksec) \\      \hline
NGC\,5408 X-1 & 0302900101	&	2006-01-13 & 130 & 99 \\
& 0500750101	&	2008-01-13 & 113 & 47   \\
& 0653380201	&	2010-07-17 & 104 & 60   \\
& 0653380301	&	2010-07-19 & 128 & 111  \\
& 0653380401	&	2011-01-26 & 119 & 90  \\
& 0653380501	&	2011-01-28 & 124 & 95  \\
& 0723130301&   2014-02-11 & 35 & 34 \\
& 0723130401&   2014-02-13 & 33 & 32  \\
NGC\,6946 X-1 & 0500730201	&	2007-11-02 & 28 & 25 \\
& 0500730101	&	2007-11-08 & 33 & 31 \\
& 0691570101	&	2012-10-21 & 114 & 99  \\
  \hline
 \end{tabular}

 \medskip
 {(a) Exposure time before the background filtering, and (b) duration after the background filtering (see text).}
\end{table}

\section[]{data reduction} \label{sample}

In this paper we utilise the multiple {\it XMM--Newton} observations of NGC\,5408 X-1 and NGC\,6946 X-1. These are among the brightest and most variable known ULXs and have some of the longest publically available {\it XMM--Newton} observations. We have in fact performed much of our analysis on all 20 ULXs in the sample discussed by \citet{sutton2013}, but for the other 18 sources the data were not sufficient to obtain meaningful constraints from the time lag and rms-flux analyses. We used {\it XMM--Newton} archival data up to March 2015.
The log of the observations used in our analyses is given in Table \ref{obs}.

\begin{figure*}
\includegraphics[width=0.35\textwidth,angle=90]{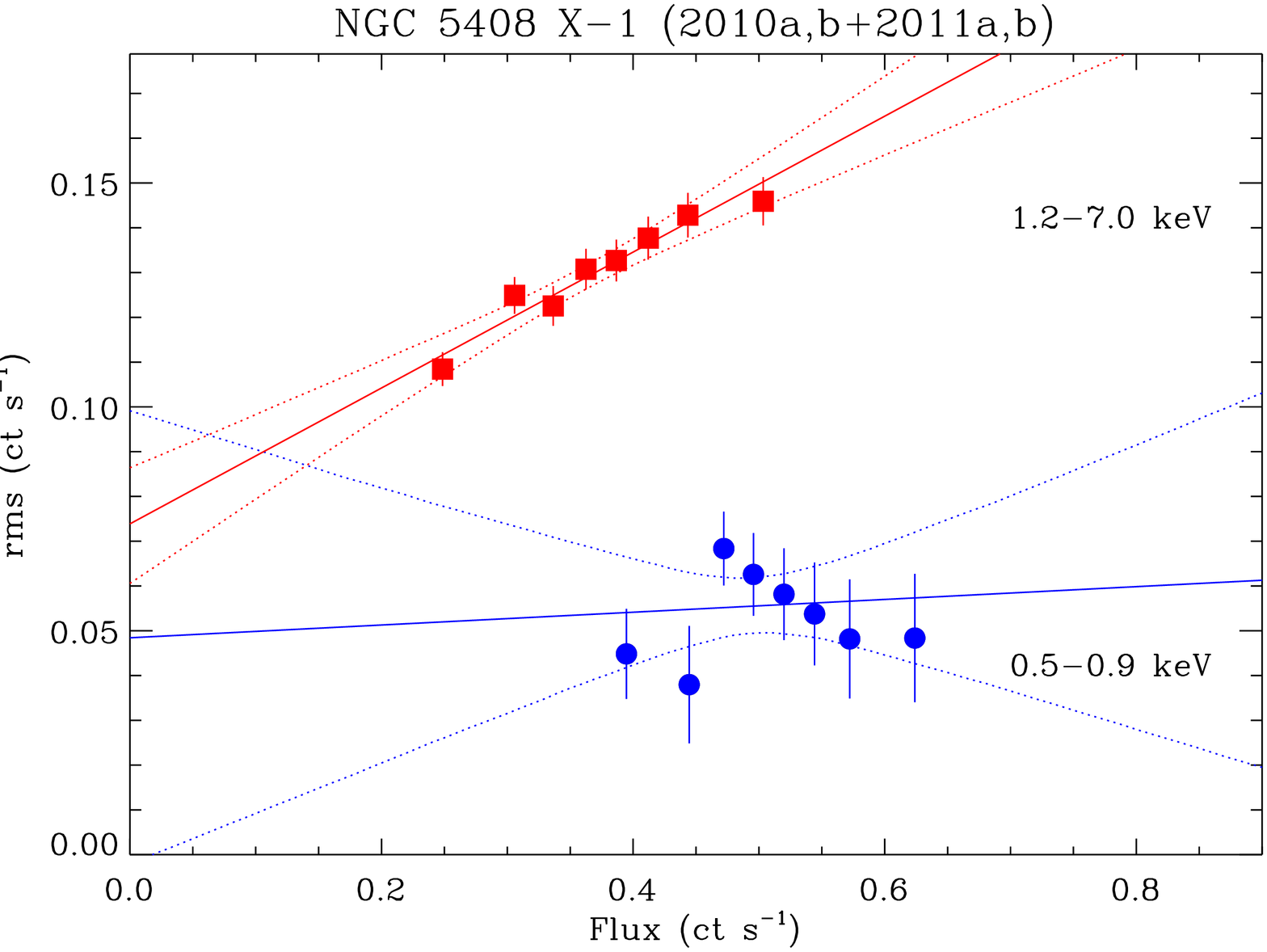}
\includegraphics[width=0.35\textwidth,angle=90]{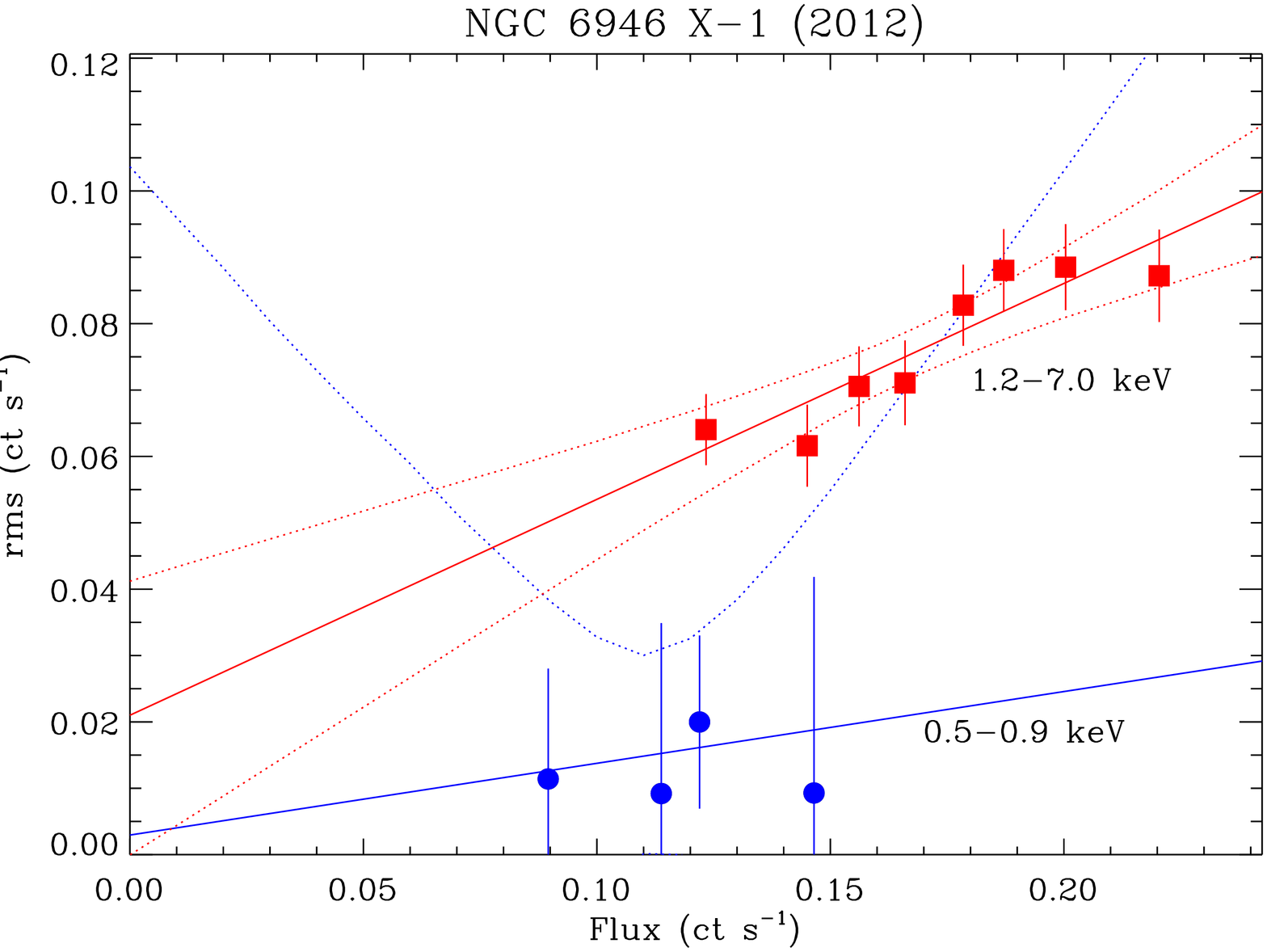}
\caption{rms-flux relation in the soft (0.5-0.9 keV) and hard (1.2-7 keV) energy bands for (left): the four observations from 2010 and 2011 of NGC\,5408 X-1, and (right): the 2012 observation of NGC\,6946 X-1. The rms is measured over the 4--50 mHz frequency range from segments of length $150$ s (NGC\,5408 X-1) and $250$ s (NGC\,6946 X-1). The dashed lines show the 95\% ``confidence bands" around the best-fitting linear model.}
\label{rmsfluxfig}
\end{figure*}

For these sources we rejected the observations with exposure times shorter than 10 ksec because the analysis was performed using continuous segments of this length in order to reach low frequencies.
To obtain a high signal-to-noise ratio (S/N) we combined data from EPIC pn, MOS1, and MOS2 detectors. We forced MOS and pn light curves to have the same {\sc tstart} and {\sc tstop}. The event files for the source were extracted from circular regions (aperture radius of 40$\arcsec$ and 25$\arcsec$ for NGC\,5408 X-1 and NGC\,6946 X-1\footnote{PN data of obsID 0200670301 and 0200670401 were extracted from elliptical regions to avoid the detector chip gaps.}) and for the background from rectangular regions, free of other sources and located in the same chip of the source, using the Science Analysis Software (SAS\footnote{http://xmm.esac.esa.int/sas/}), version 13.0.0. Standard event patterns ({\sc pattern} $\leq$ 4 for the pn detector, {\sc pattern} $\leq$ 12 for the MOS detectors) and filter ({\sc flag} = 0) were used.

The subsequent analysis was carried out using IDL software\footnote{Available from http://www.star.le.ac.uk/sav2/idl.html}.
The ULXs are relatively weak sources (in some cases $ < 0.1$ ct/s in the 2--10 keV band), and so may be overwhelmed by the background during periods of high and flaring background. Strong background flares may introduce spurious variability and time lags into the data (the spectrum of a background flare evolves with time) if not properly excluded from the data. In order to mitigate against this, we carefully filtered each observation for background flares as follows.
A light curve of the background was extracted in the 1-12 keV energy band using 2.6 s bins and smoothed using a 500 s width boxcar filter to improve the S/N. We manually selected a suitable background threshold for each observation, above which data were considered to be affected by background flares, and excluded from further analysis. 
The quiescent (non-flaring) background levels were slightly 
different between each observation. We therefore chose a slightly 
different background threshold levels for each observation in order to 
leave periods of quiescent background but effectively remove flares.
The background thresholds were in the range 0.028--0.036 ct/s for the NGC\,5408 X-1 observations, and in the range 0.010--0.015 ct/s for the NGC\,6946 X-1 observations. 

Col. 5 in Table \ref{obs} shows the total amount of ``good" exposure time after these high background periods have been excluded.
We note that the rms-flux analysis (Sect. \ref{rmsflux}) and the cross spectral analysis (Sect. \ref{cross}) were each carried out using equal length segments of continuous good data, but the lengths of the segments used for each analysis were different. This is because the rms-flux analysis concentrates on the higher frequency (i.e., shorter timescale) variations (the ``red noise" part of the PSD) while for the cross spectrum analysis we are interested in the variability properties to lower frequencies (i.e., longer timescales). Requiring fixed length segments of uninterrupted good time means some small intervals of good time were not used; the amount of ``good" data used for each analysis is therefore slightly different (lower) for each observation than the ``good" duration given in Table \ref{obs}.

Finally, we obtained an rms spectrum (defined as the square root of the normalised excess variance; see \citealt[][for details]{vaughan2003}) of the source (see Fig. \ref{rms}), and checked at which energies variability is found in order to select the energy bands.
The spectral analysis performed by \cite{middleton2011} showed the need for two spectral components to fit the spectrum of NGC\,5408 X-1, which are separated at $\sim$ 1 keV. Therefore, we will take 1 keV as the separation point for the energy band selection.
From this analysis we selected the soft and hard energy bands as 0.5--0.9 keV and  1.2--7.0 keV as these were the bands where the strongest variations were found.

\section{rms-flux relation} \label{rmsflux}

In this section we describe our analysis of the rms-flux relation in 
these two ULXs.
The analysis follows closely the analysis of \citet{heilvaughan2010} and we refer the interested reader to this paper \citep[and that of][]{heil2011} for 
specific details. Briefly, we divided each observation into continuous 
$150$ s and $250$ s segments of ``good" data, for NGC\,5408 X-1 and NGC\,6946 X-1 respectively, and from each segment computed a periodogram and the mean 
count rate. 
We needed slightly longer segments for NGC\,6946 X-1 in order to achieve a good rms estimate, since the rms is lower in this source (see Fig. \ref{rms}).
The periodograms were then averaged in eight groups 
according to the mean count rate, and the rms estimated from each. The 
rms is the square root of the Poisson noise-subtracted variance, itself 
computed by integrating the average periodograms over the $\sim 4-50$ 
mHz range, using a 10.4 s  (four times the MOS frame time) time resolution of the light curves. The analysis was performed for a soft and a hard band. Linear models of the form $\sigma = 
k(\langle F \rangle - C)$, where $F$ is the flux and $k$ (i.e., $d\sigma / d \langle F \rangle$) and $C$ are constants, were fitted to the rms-flux data using weighted least squares (min $\chi^2$). 
We also estimated the 95\% ``confidence bands" around the best-fitting linear model. We randomly generated 500 models, each one from the distribution of parameters (specified by the best-fit values and covariance matrix), and  extracted the 2.5\% and 97.5\% y values at each x value, and within these we estimated the 95\% confidence band. These illustrate the uncertainty on the fitted models.

\citet{heilvaughan2010} demonstrated a 
positive, linear rms-flux relation for the harder band data during the 
2006 and 2008 observations of NGC\,5408 X-1. Appendix \ref{rmsfluxappendix} shows the rms-flux relations for the 
2010, 2011 and 2014 observations (the best fitting linear model parameters 
were consistent between the two 2010 observations, between the two 
2011 observations, and between the two 2014 observations, and so the closely spaced pairs of observations were 
combined). In agreement with the results for 2006 and 2008 obtained by \citet{heilvaughan2010}, the 2010, 2011 and 2014 observations also show the rms-flux relation
in the harder band data, that are well fitted by a linear model (p $>$ 0.05 in all cases), 
although the parameters are slightly different between the three years. These detections (in the sense that the gradient k $>$ 0) are significant at the 4.8$\sigma$, 7.8$\sigma$ and 6.6$\sigma$ levels, respectively.
We notice that the gradients and intercepts between the 2010 and 2011 observations are consistent within the 2$\sigma$ level, so we have combined the four observations. Fig. \ref{rmsfluxfig} (left) shows the rms-flux relation for these observations of NGC\,5408 X-1, whose detection is significant at the 7.2$\sigma$ level in the hard energy band.

The soft band appears significantly different (see left panel in Fig. \ref{rmsfluxfig}, and Appendix \ref{rmsfluxappendix}); 
the gradient $k$ is much smaller. The gradient $k$ gives 
the fractional rms after subtracting any constant flux or rms components; the much lower gradient in the 
soft band indicates that, even after removing any constant component, 
the fractional amplitude of the variability is greatly suppressed. 
A linear model fits the data well (p $>$ 0.05), although all the detections are below the 3$\sigma$ level, so the rms-flux relation cannot be confirmed in the soft energy band of NGC\,5408 X-1.

We also show the rms-flux relation for NGC\,6946 X-1 for the first time (Fig. \ref{rmsfluxfig}, right panel for the 2012 observation).
Data from 2007 were used together because the linear fit was consistent between the observations  (see Appendix \ref{rmsfluxappendix}), and the 2012 data was treated separately. The rms-flux relation is clearly detected in the hard band at the 4.5$\sigma$ and 4.4$\sigma$ level for the 2007 and 2012 data, respectively. In the soft band the detections are below 3$\sigma$, although again we obtain p $>$ 0.05 in every fit with a linear model. 

Therefore, the rms-flux relation is clearly detected in the hard energy band of both ULXs, but we cannot claim this relation for the soft energy band in these ULXs.


\begin{figure}
\includegraphics[width=0.5\textwidth]{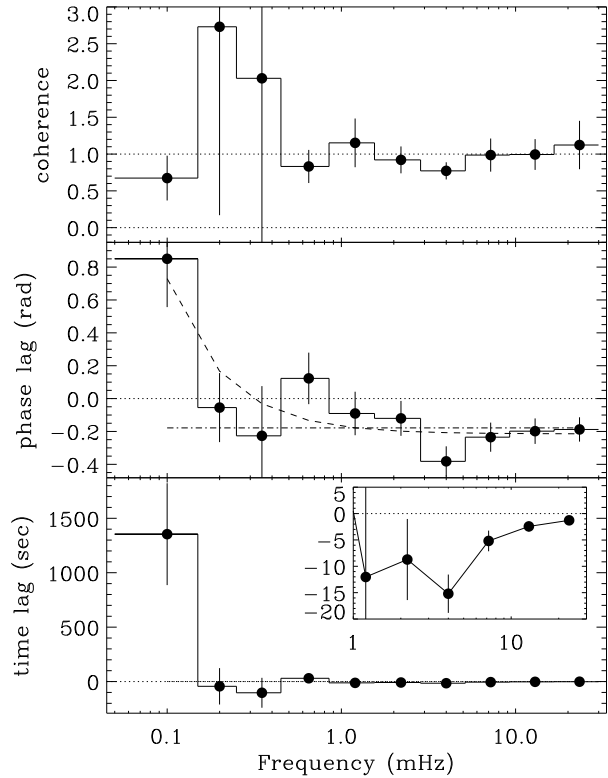}
\caption{Results of the cross-spectral analysis between the soft (0.5-0.9 keV) and hard (1.2-7.0 keV) energy bands for NGC\,5408 X-1 showing, from upper to lower panels: the coherence; phase-lag; and time lag between the energy bands (the inset panel shows a zoom-in of the soft lag). The phase lag above $\sim$1 mHz is is significantly negative, at around -0.2 rad, corresponding to a time lag of $\sim$10 s at $\sim$4 mHz. At lower frequencies the lag becomes consistent with zero or even positive lag in the lowest frequency band. The middle panel shows a constant model (dot-dashed line) and a power law plus a constant model (dashed line) fitted to the phase lag (see text).}
\label{lags}
\end{figure}

\begin{figure}
\includegraphics[width=0.34\textwidth,angle=90]{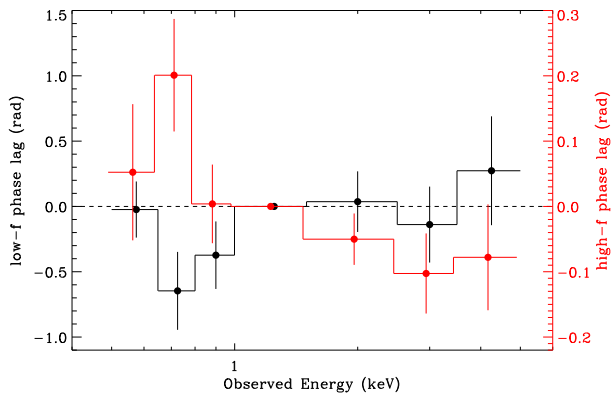}
\caption{Phase lag spectrum of NGC\,5408 X-1 at low (black; 0.1--0.3 mHz) and high (red; 1--20 mHz) frequencies. These show the average lag between a given band and the 1--1.5 keV reference band. The high frequency lag-energy spectrum decreases with energy, indicating the phase lag becomes more negative for widely separated energy bands. The y-axis are different for low (left) and high (right) frequencies.}
\label{lagspectra}
\end{figure}

\section{cross-spectrum} \label{cross}

Here we report our cross-spectral analysis, which was carried out following standard procedures \citep[see e.g.,][]{nowak1999}.
We extracted the background-subtracted source light curves in the soft (0.5-0.9 keV) and hard (1.2-7.0 keV) energy bands using time bins of 10.4 s respecting the ``good" time interval list. We used segments of 10 ks duration, and from these we estimated the power spectra (PSD), noise-subtracted coherence, phase and time lags in the Fourier frequency domain in the standard way \citep[see e.g.,][]{vaughannowak1997, nowak1999, vaughan2003}. The fractional normalisation was adopted and we rebinned over logarithmic frequency bins, each spanning a factor $\sim$1.7 in frequency.


A soft or negative lag (i.e., the soft band photons lag those in the hard band) in NGC\,5408 X-1 was first reported by \citet{heilvaughan2010} from the {\it XMM--Newton} data from 2006, and later confirmed by \citet{demarco2013b} using six {\it XMM--Newton} observations between 2006 and 2011. \citet{demarco2013b} combined data from the same year as stationarity was observed.
Since the PSDs differ between the years, as previously noted by \citet{demarco2013b}, we first undertook the analysis separately for each year, as in Sect. \ref{rmsflux}. From integrating the PSD, we found that the variability is stronger in 2010 and 2011 (four observations in total) and the phase lag spectra appear consistent, and thus we combined these data in order to obtain a higher S/N. 
The results of the cross spectrum for NGC\,5408 X-1 are shown in Fig. \ref{lags}, including the noise-subtracted coherence, phase, and time lag between the energy bands. The soft lag is detected at the 4.2$\sigma$ level. In Fig. \ref{lags} it can be seen that at lower frequencies the sign of the time lag changes and becomes positive. However, we cannot claim a hard or positive lag (i.e., the soft band photons lead those in the hard band) because this measurement is at the 2.9$\sigma$ level. The strength of the estimated lag also varies with the choice of energy bands.

Although a similar time lag behaviour is observed in NGC\,6946 X-1 (see Appendix \ref{lagsapp}), the S/N of the data is not enough to confirm the lags, whose detections are significant at the 2.3$\sigma$ (soft lag) and 1.3$\sigma$ (hard lag) level at high and low frequencies, respectively. 



We fitted two different models to the phase lag (see middle panels in Fig. \ref{lags} and Appendix \ref{lagsapp}). The first one is a constant model (dot-dashed line), and the second one consists of a power law plus a constant (dashed line) as phase lag $\Delta \phi = A f^{-\alpha} + B$, where $A$ and $B$ are constants. The second model fits both sources better, with $A = 0.05 \pm 0.07$, $B = -0.21 \pm 0.04$, $\alpha = 1.3 \pm 0.7$, $\chi^2/d.o.f = 9.2/7$ and $p=0.24$ for NGC\,5408 X-1, and $A = 0.7 \pm 0.7$, $B = -0.6 \pm 0.6$, $\alpha = 0.4 \pm 0.4$, $\chi^2/d.o.f = 1.9/7$, and $p=0.97$ for NGC\,6946 X-1. 
With this model the hard lag at low frequencies has a $\Delta \phi \propto f^{-1}$ dependence, and a constant phase soft lag of -0.2 rad at high frequencies.
Therefore the phase lag is not constant with frequency, nor is the time lag.

For NGC\,5408 X-1 we also computed the phase lag vs. energy spectrum from the cross-spectra, averaging over the frequency bands.                
Fig. \ref{lagspectra} shows the phase lag spectrum from the 2010 and 2011 observations of NGC\,5408 X-1 at low (0.1-0.3 mHz, in black) and high (1-20 mHz, in red) frequencies, corresponding to the hard and soft lags for NGC\,5408 X-1.
The 1.0-1.5 keV energy band was used as the reference band because this is where the signal is highest, i.e, the rms is high.
The lags have not been shifted, so zero-lag means that there is no time delay
between that bin and the reference band. Similarly, a negative lag means that the bin leads the reference band, and positive lags mean that the bin lags behind the reference band.
At low frequencies (i.e., the hard lag) the phase lag increases with energy, from $\sim$ -0.6 at 0.8 keV to $\sim$ 0.3 at 3 keV.
Conversely, at high frequencies (i.e., the soft lag) the phase lag spectrum decreases with energy, from $\sim$ 0.2 at 0.8 keV down to $\sim$ -0.1 at 3 keV. Note that the y-axis is different for low (left y-axis) and high (right y-axis) frequencies. The same behaviour was reported by \citet{demarco2013b} for the time lag vs. energy spectrum of the soft lag.

\section[]{Discussion}  \label{discussion}

We have shown that a linear rms-flux relation is present in the hard energy band of the two ULXs studied here, NGC\,5408 X-1 and NGC\,6946 X-1, and that it is persistent across multiple observations spanning $\sim$8 years. The limitations of the soft band data are such that we are unable to place any interesting constraints on the rms-flux relation in the soft band (this is due to the relatively low total count rate of these sources, $\lesssim 1$ ct s$^{-1}$, and the low fractional rms amplitude below 1 keV, which means the intrinsic rms is low, the Poisson noise is relatively high and the flux range is small). We also examined the frequency-dependent X-ray time lags, extending the analysis to lower frequencies than previous studies. We find the coherence between soft ($<$0.9 keV) and hard ($>$1.2 keV) bands is high, consistent with $\approx 1$ at all frequencies. The soft phase lag can be recovered for most observations of NGC\,5408 X-1 at frequencies above $\sim$few mHz, but the phase lag is not constant down to the lowest frequencies we probe ($\sim$0.1 mHz) and becomes less negative at lower frequencies. We find tentative evidence for a hard lag (positive phase lag) at the lowest frequencies.

\subsection{The rms-flux relation}

\cite{heilvaughan2010} used the 2006 and 2008 observations to show that NGC\,5408 X-1 exhibits a linear rms-flux relation. We show the relation persists through further observations in 2010, 2011 and 2014. The parameters of the best-fitting linear models differ between observations, presumably due to modest changes in mean flux and PSD shape. We also detected a linear rms-flux relation in the 2012 observation of NGC\,6946 X-1, demonstrating that NGC\,5408 X-1 is not unique among ULXs. 
We did examine the X-ray variability of the other ULXs from the sample 
of \citet{sutton2013} but we were unable to obtain useful constraints on 
the possible rms-flux relations. Given that we did detect the linear 
rms-flux relation in the two best observed, highly variable ULXs, we 
speculate that the rms-flux relation is common to variable ULXs but that 
in most other cases the low count rates and shorter observations 
preclude a good detection.


The rms-flux relation appears to be ubiquitous in luminous, accreting objects, such as active galactic nuclei \citep[AGN,][]{bvaughan2003,vaughan2003}, Galactic X-ray binaries (XRB) with black hole or neutron star accretors \citep{uttley2001,uttley2004} and cataclysmic variables \citep{scaringi2012,vandesande2015}. The simplest conclusion is that a common physical mechanism explains the strong, aperiodic variability, following a linear rms-flux relation, in all these sources. Currently the favoured explanation is in terms of propagating accretion rate fluctuations \citep{lyubarskii1997, arevalo2006}, although any similar scheme for multiplicative modulation of random fluctuations will produce a linear rms-flux relation \citep{uttley2005}. In this model, random fluctuations in the viscosity occurring on all spatial scales in the accretion flow modulate the accretion rate further in, but high frequency variations occurring on large scales are damped as they move inwards. Irrespective of this, the presence of the linear rms-flux relation in more than one ULX suggests the same underlying variability mechanism is at work in ULXs as in the sub-Eddington XRB and AGN.

\subsection{Interband X-ray time delays}

The soft lag in NGC\,5408 X-1 was first detected by \cite{heilvaughan2010}, based mainly on the 2006 {\it XMM--Newton} observation of NGC\,5408 X-1. \cite{demarco2013b} subsequently used six {\it XMM--Newton} observations (2006, 2008, 2010a,b and 2011a,b) and recovered similar soft X-ray lags in each, above frequencies of $\sim$few mHz. We extended the lag-frequency analysis to lower frequencies ($\sim 0.1$ mHz) and find that the soft lag extends down to $\sim 1$ mHz with an amplitude of tens of seconds, but neither the phase lag nor time lag is constant with frequency, with the phase lag constant at $\sim -0.2$ rad (soft lag) above $\sim$1~mHz but increasing at lower frequencies. A similar lag-frequency spectrum is found for NGC\,6946 X-1.

The soft vs. hard band coherence is high, indicating the variations in each band are linearly correlated, and probably driven by the same process. Any additional source of uncorrelated variations adding to one band must be weak. The lags may then be imposed by the action of linear filters (``impulse response functions'') acting on the driving variations in one or both bands.

The interpretation of these lags remains challenging. In the following, we discuss possible implications and comparison with other accreting sources.

\subsubsection{\label{agn}Comparison with AGN scaling relations}

Soft X-ray time lags are now well established in AGN \citep[see][for a review]{uttley2014}, where they are observed typically at frequencies above $\sim 0.1$ mHz, below which a hard lag dominates. The most accepted scenario for AGN assumes that the hard lag is produced by accretion rate fluctuations moving inwards through the accretion disc and energising a radially extended corona \citep[e.g.,][]{kotov2001, arevalo2006}. The soft lags are thought to be produced by a separate process -- ``reverberation'', i.e.  light-travel time delays as the primary coronal emission is reprocessed in the inner regions of a partially ionised disc (see \citealt{fabian2009}, and the review by \citealt{uttley2014}, and references therein). Typically the reverberation is thought to be generated by the inner disc, within $\sim 10 r_g$ of the central black hole.

\cite{demarco2013a} studied a sample of 32 AGN and found a scaling relation between the black hole mass and the soft time lag using the 15 objects where a soft time delay was detected. If their scaling relation extends to ULXs, we would expect the soft lag to be located at frequencies in the range $20-200$ mHz with amplitudes of $0.03-0.4$ s for black hole masses of $10-1000M_{\odot}$, as expected for stellar-mass to intermediate-mass black holes \citep[see e.g. Fig. 9 of][]{demarco2013b}. As shown in Fig. \ref{lags}, the soft lag in NGC\,5408 X-1 occurs at frequencies of the order of mHz with amplitudes of tens of seconds, much larger than expected for a stellar mass black hole binary and more typical of a low-mass AGN.

However, NGC\,5408 X-1 is clearly identified with a stellar counterpart and
associated nebulosity on the outskirts of the dwarf galaxy NGC\,5408 and is therefore
highly unlikely to be an AGN \citep[e.g.,][]{pakull2003,cseh2013}.
Indeed, its X-ray spectrum and coarse timing properties fit very nearly into the
range of known behaviours for ULXs in the proposed `ultraluminous state' \citep{gladstone2009, middleton2011, sutton2013}, indicating it is likely to
be a relatively small black hole accreting at super-Eddington rates \citep{motch2014}.





\subsubsection{Association with XRB QPOs}

Another suggestion discussed by \cite{demarco2013b} is that the soft lag is associated with a QPO, as sometimes occurs in BH XRBs. There are some problems with this interpretation. Foremost is that the QPOs claimed for NGC\,5408 X-1 are generally not clear and distinct features like QPOs in BH XRBs (of the ``type-C'' variety). We will discuss this further in a companion paper (Vaughan et al., in prep.), but even if we take the PSD fits of \cite{pasham2012}, \cite{demarco2013b} or \cite{caballerogarcia2013} at face value, the soft lags extend over a much broader range of frequencies than the QPO-like features, indicating the lags are associated with the noise PSD. It is also not clear that lags in ULXs, e.g., between the 0.5--0.9 keV and 1.2--7 keV energy bands at $\sim$1 mHz QPOs, can be compared to those in XRBs, which are usually observed between harder energy bands (e.g., 3--5 keV and 5--13 keV) for QPOs at $\sim$1 Hz \citep{remillard2002,casella2004}.

\subsubsection{Possible origins for the soft lag}

The most popular explanation of soft X-ray time lags in AGN is in terms 
of reverberation \citep[e.g.,][]{fabian2009, uttley2014}. Can a 
similar explanation hold in these ULXs? \cite{caballerogarcia2010} 
fit the {\it XMM--Newton} spectrum of NGC\,5408 X-1 using reflection models. Their 
best-fitting model includes reflection from an ionised disc extending in 
to $\sim 2.5 r_g$. Their model accounts for the ``soft excess'' in terms 
of emission from the disc (e.g., ionised N, O, Fe-L emission lines and 
recombination continua), rather than using an additional continuum 
component, and so represents an upper limit on the strength of a 
reflection component.

This model does not easily explain the timing properties. If the soft 
lag is the result of strong soft X-ray reflection responding to the 
primary X-ray luminosity, then the soft excess should be present in the 
covariance spectrum, but the observed covariance spectra for NGC\,5408 
X-1 and NGC\,6946 X-1 both show no soft excess \citep{middleton2015}. 
Furthermore, the magnitude of the delay would imply an unreasonably 
large $M_{BH}$ (see Sect. \ref{agn}). Most of the reverberation signal in the reflection model 
is produced in the inner $<10 r_g$. If the observed maximum soft time 
lag of $\sim 10$ s is not ``diluted'' (due to the contribution of direct 
emission to both bands), then the delay corresponds to a path length of 
$\sim 3 \times 10^6$ km. This implies a distance of $\sim 2 \times 10^4 (M_{BH}/100 M_{\odot}) r_g$ from the illuminating X-ray source. Unless the 
black hole mass is $\gsim 10^5 M_{\odot}$, comparable to a low mass AGN, the 
reverberation must be originating at $\gg 10^2 r_g$, inconsistent with the 
reflection model of \cite{caballerogarcia2010}. Since the effect of 
dilution is to make the reflection site even more distant in order to 
produce the observed delay, it  would most likely produce strong 
 emissions lines (e.g., Fe K$\alpha$) that are relatively narrow, which 
 are not seen in ULX spectra \citep[e.g.,][]{gladstone2009,sutton2013}.

Indeed, recent {\it NuSTAR\/} observations \citep[e.g.,][]{bachetti2013, walton2014, walton2015a, walton2015b} have demonstrated that ULX spectra lack the obvious signatures of strong
reflection (such as iron emission lines and the ``Compton hump'') commonly seen in Seyfert 1s
that display a soft lag reverberation signal \citep[e.g.,][]{zoghbi2011, kara2013}. 
In particular, \cite{walton2015a} studied {\it NuSTAR, XMM-Newton} and {\it Suzaku} observations of Holmberg\,II X-1, this ULX being the closest analogue to NGC\,5408 X-1 and NGC\,6946 X-1 that has been observed by {\it NuSTAR} \citep{caballerogarcia2010,sutton2013}, and found no evidence of reflection in the spectrum.
Hence a similar reverberation origin for the soft lags in ULXs appears unlikely, as reflection is not a strong component of the typical ULX spectra.
We note that even if {\it NuSTAR\/} data would be useful to rule out (or not) the reverberation origin of the soft lag, NGC\,5408 X-1 is too soft to make a good {\it NuSTAR\/} target.

The spectrum of these ``soft ultraluminous'' sources such as NGC\,5408 X-1 and NGC\,6946 X-1 are often described in terms of two components  \citep{gladstone2009,sutton2013,middleton2015}. The softer component, dominating below $\sim 1$ keV has a quasi-thermal spectrum and shows little, if any, short timescale variability. The latter fact is inferred from the lack of the ``soft excess'' in the rms and covariance spectra (see Fig. \ref{rms} and also \citealt{middleton2015}). The harder component, dominating over $\sim 1-10$ keV, resembles a cut-off power law \citep[see also][]{walton2013,walton2014}, and produces most or all of the rapid variability. In the context of the supercritical accretion model \citep{shakura1973,king2001,poutanen2007,middleton2015} the hard component represents emission from the inner accretion flow, while the soft component is thermal emission from the base of an optically thick, massive wind driven off the disc at larger radii.

The lack of variability of the softer component suggests it plays no role in generating the soft lags. Further, the high coherence between soft and hard bands is most simply explained if there is a single ``driver'' of the variability. Together, these are consistent with the soft lag being intrinsic to the harder component and not a delay between the two spectral components. In other words, it is only the hard component that varies rapidly, and its variations on timescales shorter than $\sim 1$ ks occur at higher energies first, with the softer emission (from the same spectral component) taking up to $\sim$ few seconds to respond. These delays may be intrinsic to the emission mechanism (from the inner accretion flow) or imposed by processes intercepting and delaying some fraction of this emission.

One explanation for X-ray time lags is in terms of scattering in an intervening medium. A soft lag could be produced as hard X-ray photons from the primary X-ray source pass through and are down-scattered to lower energies. Low energy photons typically have undergone more scatterings and so escape the scattering medium after a longer delay. The lag-frequency spectrum for NGC\,5408 X-1 shows clear frequency dependence, being constant in neither phase lag nor time lag. Such lags are however difficult to reproduce in a simple scattering scenario which would more naturally produce an approximately constant time lag \citep[at frequencies below the “wrap around” frequency;][]{miller2010,zoghbi2011,uttley2014}.

A further possibility is that the soft lag somehow results from the propagation of photons through an optically thin shroud of material, likely to result from an expanded wind (a prediction of the super-critical ULX model). Should the absorption opacity of this material be low in the hard band and high in the soft band (perhaps as a result of high abundances of Oxygen and Neon but relatively low abundances of Iron, \citealt{middleton2014}), the hard photons will arrive to the observer scattered (with the exact scattered fraction dependent on the Thompson optical depth) whilst the soft photons will be absorbed and re-emitted. There are indeed strong indications that residuals at soft energies in the time-averaged spectra are associated with absorption and emission features associated with a strong outflow (\citealt{middleton2014}; Middleton et al. in prep) and, should this provide an origin for the soft lag, the magnitude of the lag should provide constraints on the extent of the wind. However, such models incorporating outflows are by their very nature ``messy'' with the impact of reverberation and absorption heavily dependent on inclination angle and structure of the wind.

A final possibility is that the lags are intrinsic to the X-ray source itself, i.e., that the part of the inner accretion flow that first responds to inward moving accretion rate variations has a harder spectrum than the later responding parts of the flow. This is in the opposite sense to the models used to explain hard lags in XRBs \citep[e.g.,][]{kotov2001,arevalo2006}. But we should perhaps not rule out this idea, as the structure of the hard X-ray emitting inner regions of ULXs may be quite different from the corona or jet-base thought to produce the thermal/non-thermal, variable hard X-ray spectrum in sub-Eddington XRBs and AGN.

\subsubsection{Low frequency lags}

At frequencies below $\sim$ 1 mHz the sign of the time lag estimate changes and becomes a hard (or positive) lag. We do caution that, even with multiple observations of one of the brightest ULXs, the lag at $\sim 0.1$ mHz has large uncertainties, with the lag exceeding zero only at the $2-3 \sigma$ level.

Hard lags at low frequencies, along with soft lags at higher frequencies, have been observed in many AGN \citep{fabian2009,zoghbi2011,emmanoulopoulos2011,kara2013,demarco2013a}. The best studied source is the narrow line Seyfert 1 1H0707-495 \citep[e.g.,][]{fabian2009,zoghbi2011}. \cite{kara2013} argued that the different energy dependence of the high and low frequency lags (soft and hard, respectively) in this object revealed that different emission processes are involved. Interestingly, the lag-energy spectra for NGC\,5408 X-1 at low and high frequencies (Fig. \ref{lagspectra}) look like very similar (after a change of sign and scaling). But given the low significance of the low frequency hard lag it would perhaps be premature to place much weight on this.

\hspace*{1cm}

The limited spectral-timing data for ULXs leaves few clues about the origin of the lags. Longer observations with higher S/N would be able to access both lower and higher frequencies, and allow us to better estimate the frequency and energy dependence of the lags. Such constraints will be crucial for ruling out lag models, but may require future, larger X-ray missions.

\section*{Acknowledgments}

We thank the anonymous referee for her/his helpful comments.
This paper is based on observations obtained with {\it XMM--Newton}, an ESA science mission with instruments and contributions directly funded by ESA Member States and the USA (NASA). This research has made use of NASA's Astrophysics Data System and of data, software and web tools obtained from NASA's High Energy Astrophysics Science Archive Research Center (HEASARC), a service of Goddard Space Flight Center and the Smithsonian Astrophysical Observatory. This work was financed by MINECO grant AYA 2010-15169 and AYA 2013-42227-P. SV, TPR are supported in part by STFC consolidated grants ST/K001000/1 and ST/L00075X/1, respectively, MJM appreciates support via ERC grant 340442, and LHG acknowledges financial support from the Ministerio de Econom\'{i}a y Competitividad through the Spanish grants FPI BES-2011-043319 and EEBB-I-14-07885.

\bibliographystyle{mn2e}
\bibliography{ULX}

\begin{thebibliography}{}
\makeatletter
\relax
\def\mn@urlcharsother{\let\do\@makeother \do\$\do\&\do\#\do\^\do\_\do\%\do\~}
\def\mn@doi{\begingroup\mn@urlcharsother \@ifnextchar[{\mn@doi@}{\mn@doi@[]}}
\def\mn@doi@[#1]#2{\def\@tempa{#1}\ifx\@tempa\@empty
  \href{http://dx.doi.org/#2}{doi:#2}\else \href{http://dx.doi.org/#2}{#1}\fi
  \endgroup}
\def\mn@eprint#1#2{\mn@eprint@#1:#2::\@nil}
\def\mn@eprint@arXiv#1{\href{http://arxiv.org/abs/#1}{{\tt arXiv:#1}}}
\def\mn@eprint@dblp#1{\href{http://dblp.uni-trier.de/rec/bibtex/#1.xml}{dblp:#%
1}}
\def\mn@eprint@#1:#2:#3:#4\@nil{\def\@tempa {#1}\def\@tempb {#2}\def\@tempc
  {#3}\ifx \@tempc \@empty \let\@tempc\@tempb \let\@tempb\@tempa \fi \ifx
  \@tempb \@empty \def\@tempb{arXiv}\fi \@ifundefined
  {mn@eprint@\@tempb}{\@tempb:\@tempc}{\expandafter \expandafter \csname
  mn@eprint@\@tempb\endcsname \expandafter{\@tempc}}}

\bibitem[\protect\citeauthoryear{{Ar{\'e}valo} \& {Uttley}}{{Ar{\'e}valo} \&
  {Uttley}}{2006}]{arevalo2006}
{Ar{\'e}valo} P.,  {Uttley} P.,  2006, \mn@doi [\mnras]
  {10.1111/j.1365-2966.2006.09989.x}, \href
  {http://adsabs.harvard.edu/abs/2006MNRAS.367..801A} {367, 801}

\bibitem[\protect\citeauthoryear{{Bachetti} et~al.,}{{Bachetti}
  et~al.}{2013}]{bachetti2013}
{Bachetti} M.,  et~al., 2013, \mn@doi [\apj] {10.1088/0004-637X/778/2/163},
  \href {http://adsabs.harvard.edu/abs/2013ApJ...778..163B} {778, 163}

\bibitem[\protect\citeauthoryear{{Bachetti} et~al.,}{{Bachetti}
  et~al.}{2014}]{bachetti2014}
{Bachetti} M.,  et~al., 2014, \mn@doi [\nat] {10.1038/nature13791}, \href
  {http://adsabs.harvard.edu/abs/2014Natur.514..202B} {514, 202}

\bibitem[\protect\citeauthoryear{{Caballero-Garc{\'i}a} \&
  {Fabian}}{{Caballero-Garc{\'i}a} \& {Fabian}}{2010}]{caballerogarcia2010}
{Caballero-Garc{\'i}a} M.~D.,  {Fabian} A.~C.,  2010, \mn@doi [\mnras]
  {10.1111/j.1365-2966.2009.16062.x}, \href
  {http://adsabs.harvard.edu/abs/2010MNRAS.402.2559C} {402, 2559}

\bibitem[\protect\citeauthoryear{{Caballero-Garc{\'i}a}, {Belloni}  \&
  {Wolter}}{{Caballero-Garc{\'i}a} et~al.}{2013}]{caballerogarcia2013}
{Caballero-Garc{\'i}a} M.~D.,  {Belloni} T.~M.,   {Wolter} A.,  2013, \mn@doi
  [\mnras] {10.1093/mnras/stt1479}, \href
  {http://adsabs.harvard.edu/abs/2013MNRAS.435.2665C} {435, 2665}

\bibitem[\protect\citeauthoryear{{Casella}, {Belloni}, {Homan}  \&
  {Stella}}{{Casella} et~al.}{2004}]{casella2004}
{Casella} P.,  {Belloni} T.,  {Homan} J.,   {Stella} L.,  2004, \mn@doi [\aap]
  {10.1051/0004-6361:20041231}, \href
  {http://adsabs.harvard.edu/abs/2004A\%26A...426..587C} {426, 587}

\bibitem[\protect\citeauthoryear{{Cseh}, {Gris{\'e}}, {Kaaret}, {Corbel},
  {Scaringi}, {Groot}, {Falcke}  \& {K{\"o}rding}}{{Cseh}
  et~al.}{2013}]{cseh2013}
{Cseh} D.,  {Gris{\'e}} F.,  {Kaaret} P.,  {Corbel} S.,  {Scaringi} S.,
  {Groot} P.,  {Falcke} H.,   {K{\"o}rding} E.,  2013, \mn@doi [\mnras]
  {10.1093/mnras/stt1484}, \href
  {http://adsabs.harvard.edu/abs/2013MNRAS.435.2896C} {435, 2896}

\bibitem[\protect\citeauthoryear{{De Marco}, {Ponti}, {Cappi}, {Dadina},
  {Uttley}, {Cackett}, {Fabian}  \& {Miniutti}}{{De Marco}
  et~al.}{2013a}]{demarco2013a}
{De Marco} B.,  {Ponti} G.,  {Cappi} M.,  {Dadina} M.,  {Uttley} P.,  {Cackett}
  E.~M.,  {Fabian} A.~C.,   {Miniutti} G.,  2013a, \mn@doi [\mnras]
  {10.1093/mnras/stt339}, \href
  {http://adsabs.harvard.edu/abs/2013MNRAS.431.2441D} {431, 2441}

\bibitem[\protect\citeauthoryear{{De Marco}, {Ponti}, {Miniutti}, {Belloni},
  {Cappi}, {Dadina}  \& {Mu{\~n}oz-Darias}}{{De Marco}
  et~al.}{2013b}]{demarco2013b}
{De Marco} B.,  {Ponti} G.,  {Miniutti} G.,  {Belloni} T.,  {Cappi} M.,
  {Dadina} M.,   {Mu{\~n}oz-Darias} T.,  2013b, \mn@doi [\mnras]
  {10.1093/mnras/stt1853}, \href
  {http://adsabs.harvard.edu/abs/2013MNRAS.436.3782D} {436, 3782}

\bibitem[\protect\citeauthoryear{{Dheeraj} \& {Strohmayer}}{{Dheeraj} \&
  {Strohmayer}}{2012}]{pasham2012}
{Dheeraj} P.~R.,  {Strohmayer} T.~E.,  2012, \mn@doi [\apj]
  {10.1088/0004-637X/753/2/139}, \href
  {http://adsabs.harvard.edu/abs/2012ApJ...753..139D} {753, 139}

\bibitem[\protect\citeauthoryear{{Emmanoulopoulos}, {McHardy}  \&
  {Papadakis}}{{Emmanoulopoulos} et~al.}{2011}]{emmanoulopoulos2011}
{Emmanoulopoulos} D.,  {McHardy} I.~M.,   {Papadakis} I.~E.,  2011, \mn@doi
  [\mnras] {10.1111/j.1745-3933.2011.01106.x}, \href
  {http://adsabs.harvard.edu/abs/2011MNRAS.416L..94E} {416, L94}

\bibitem[\protect\citeauthoryear{{Fabian} et~al.,}{{Fabian}
  et~al.}{2009}]{fabian2009}
{Fabian} A.~C.,  et~al., 2009, \mn@doi [\nat] {10.1038/nature08007}, \href
  {http://adsabs.harvard.edu/abs/2009Natur.459..540F} {459, 540}

\bibitem[\protect\citeauthoryear{{Farrell}, {Webb}, {Barret}, {Godet}  \&
  {Rodrigues}}{{Farrell} et~al.}{2009}]{farrell2009}
{Farrell} S.~A.,  {Webb} N.~A.,  {Barret} D.,  {Godet} O.,   {Rodrigues} J.~M.,
   2009, \mn@doi [\nat] {10.1038/nature08083}, \href
  {http://adsabs.harvard.edu/abs/2009Natur.460...73F} {460, 73}

\bibitem[\protect\citeauthoryear{{Feng} \& {Soria}}{{Feng} \&
  {Soria}}{2011}]{feng2011}
{Feng} H.,  {Soria} R.,  2011, \mn@doi [\nar] {10.1016/j.newar.2011.08.002},
  \href {http://adsabs.harvard.edu/abs/2011NewAR..55..166F} {55, 166}

\bibitem[\protect\citeauthoryear{{Gladstone}, {Roberts}  \& {Done}}{{Gladstone}
  et~al.}{2009}]{gladstone2009}
{Gladstone} J.~C.,  {Roberts} T.~P.,   {Done} C.,  2009, \mn@doi [\mnras]
  {10.1111/j.1365-2966.2009.15123.x}, \href
  {http://adsabs.harvard.edu/abs/2009MNRAS.397.1836G} {397, 1836}

\bibitem[\protect\citeauthoryear{{Heil} \& {Vaughan}}{{Heil} \&
  {Vaughan}}{2010}]{heilvaughan2010}
{Heil} L.~M.,  {Vaughan} S.,  2010, \mn@doi [\mnras]
  {10.1111/j.1745-3933.2010.00864.x}, \href
  {http://adsabs.harvard.edu/abs/2010MNRAS.405L..86H} {405, L86}

\bibitem[\protect\citeauthoryear{{Heil}, {Vaughan}  \& {Roberts}}{{Heil}
  et~al.}{2009}]{heilvaughan2009}
{Heil} L.~M.,  {Vaughan} S.,   {Roberts} T.~P.,  2009, \mn@doi [\mnras]
  {10.1111/j.1365-2966.2009.15068.x}, \href
  {http://adsabs.harvard.edu/abs/2009MNRAS.397.1061H} {397, 1061}

\bibitem[\protect\citeauthoryear{{Heil}, {Vaughan}  \& {Uttley}}{{Heil}
  et~al.}{2011}]{heil2011}
{Heil} L.~M.,  {Vaughan} S.,   {Uttley} P.,  2011, \mn@doi [\mnras]
  {10.1111/j.1745-3933.2010.00997.x}, \href
  {http://adsabs.harvard.edu/abs/2011MNRAS.411L..66H} {411, L66}

\bibitem[\protect\citeauthoryear{{Heil}, {Vaughan}  \& {Uttley}}{{Heil}
  et~al.}{2012}]{heil2012}
{Heil} L.~M.,  {Vaughan} S.,   {Uttley} P.,  2012, \mn@doi [\mnras]
  {10.1111/j.1365-2966.2012.20824.x}, \href
  {http://adsabs.harvard.edu/abs/2012MNRAS.422.2620H} {422, 2620}

\bibitem[\protect\citeauthoryear{{Kara}, {Fabian}, {Cackett}, {Steiner},
  {Uttley}, {Wilkins}  \& {Zoghbi}}{{Kara} et~al.}{2013}]{kara2013}
{Kara} E.,  {Fabian} A.~C.,  {Cackett} E.~M.,  {Steiner} J.~F.,  {Uttley} P.,
  {Wilkins} D.~R.,   {Zoghbi} A.,  2013, \mn@doi [\mnras]
  {10.1093/mnras/sts155}, \href
  {http://adsabs.harvard.edu/abs/2013MNRAS.428.2795K} {428, 2795}

\bibitem[\protect\citeauthoryear{{King}}{{King}}{2009}]{king2009}
{King} A.~R.,  2009, \mn@doi [\mnras] {10.1111/j.1745-3933.2008.00594.x}, \href
  {http://adsabs.harvard.edu/abs/2009MNRAS.393L..41K} {393, L41}

\bibitem[\protect\citeauthoryear{{King}, {Davies}, {Ward}, {Fabbiano}  \&
  {Elvis}}{{King} et~al.}{2001}]{king2001}
{King} A.~R.,  {Davies} M.~B.,  {Ward} M.~J.,  {Fabbiano} G.,   {Elvis} M.,
  2001, \mn@doi [\apjl] {10.1086/320343}, \href
  {http://adsabs.harvard.edu/abs/2001ApJ...552L.109K} {552, L109}

\bibitem[\protect\citeauthoryear{{Kotov}, {Churazov}  \& {Gilfanov}}{{Kotov}
  et~al.}{2001}]{kotov2001}
{Kotov} O.,  {Churazov} E.,   {Gilfanov} M.,  2001, \mn@doi [\mnras]
  {10.1046/j.1365-8711.2001.04769.x}, \href
  {http://adsabs.harvard.edu/abs/2001MNRAS.327..799K} {327, 799}

\bibitem[\protect\citeauthoryear{{Liu}, {Bregman}, {Bai}, {Justham}  \&
  {Crowther}}{{Liu} et~al.}{2013}]{liu2013}
{Liu} J.-F.,  {Bregman} J.~N.,  {Bai} Y.,  {Justham} S.,   {Crowther} P.,
  2013, \mn@doi [\nat] {10.1038/nature12762}, \href
  {http://adsabs.harvard.edu/abs/2013Natur.503..500L} {503, 500}

\bibitem[\protect\citeauthoryear{{Lyubarskii}}{{Lyubarskii}}{1997}]{lyubarskii%
1997}
{Lyubarskii} Y.~E.,  1997, \mnras, \href
  {http://adsabs.harvard.edu/abs/1997MNRAS.292..679L} {292, 679}

\bibitem[\protect\citeauthoryear{{Mezcua}, {Roberts}, {Lobanov}  \&
  {Sutton}}{{Mezcua} et~al.}{2015}]{mezcua2015}
{Mezcua} M.,  {Roberts} T.~P.,  {Lobanov} A.~P.,   {Sutton} A.~D.,  2015,
  \mn@doi [\mnras] {10.1093/mnras/stv143}, \href
  {http://adsabs.harvard.edu/abs/2015MNRAS.448.1893M} {448, 1893}

\bibitem[\protect\citeauthoryear{{Middleton}, {Roberts}, {Done}  \&
  {Jackson}}{{Middleton} et~al.}{2011}]{middleton2011}
{Middleton} M.~J.,  {Roberts} T.~P.,  {Done} C.,   {Jackson} F.~E.,  2011,
  \mn@doi [\mnras] {10.1111/j.1365-2966.2010.17712.x}, \href
  {http://adsabs.harvard.edu/abs/2011MNRAS.411..644M} {411, 644}

\bibitem[\protect\citeauthoryear{{Middleton} et~al.,}{{Middleton}
  et~al.}{2013}]{middleton2013}
{Middleton} M.~J.,  et~al., 2013, \mn@doi [\nat] {10.1038/nature11697}, \href
  {http://adsabs.harvard.edu/abs/2013Natur.493..187M} {493, 187}

\bibitem[\protect\citeauthoryear{{Middleton}, {Walton}, {Roberts}  \&
  {Heil}}{{Middleton} et~al.}{2014}]{middleton2014}
{Middleton} M.~J.,  {Walton} D.~J.,  {Roberts} T.~P.,   {Heil} L.,  2014,
  \mn@doi [\mnras] {10.1093/mnrasl/slt157}, \href
  {http://adsabs.harvard.edu/abs/2014MNRAS.438L..51M} {438, L51}

\bibitem[\protect\citeauthoryear{{Middleton}, {Heil}, {Pintore}, {Walton}  \&
  {Roberts}}{{Middleton} et~al.}{2015}]{middleton2015}
{Middleton} M.~J.,  {Heil} L.,  {Pintore} F.,  {Walton} D.~J.,   {Roberts}
  T.~P.,  2015, \mn@doi [\mnras] {10.1093/mnras/stu2644}, \href
  {http://adsabs.harvard.edu/abs/2015MNRAS.447.3243M} {447, 3243}

\bibitem[\protect\citeauthoryear{{Miller}, {Turner}, {Reeves}  \&
  {Braito}}{{Miller} et~al.}{2010}]{miller2010}
{Miller} L.,  {Turner} T.~J.,  {Reeves} J.~N.,   {Braito} V.,  2010, \mn@doi
  [\mnras] {10.1111/j.1365-2966.2010.17261.x}, \href
  {http://adsabs.harvard.edu/abs/2010MNRAS.408.1928M} {408, 1928}

\bibitem[\protect\citeauthoryear{{Motch}, {Pakull}, {Soria}, {Gris{\'e}}  \&
  {Pietrzy{\'n}ski}}{{Motch} et~al.}{2014}]{motch2014}
{Motch} C.,  {Pakull} M.~W.,  {Soria} R.,  {Gris{\'e}} F.,   {Pietrzy{\'n}ski}
  G.,  2014, \mn@doi [\nat] {10.1038/nature13730}, \href
  {http://adsabs.harvard.edu/abs/2014Natur.514..198M} {514, 198}

\bibitem[\protect\citeauthoryear{{Nowak}, {Vaughan}, {Wilms}, {Dove}  \&
  {Begelman}}{{Nowak} et~al.}{1999}]{nowak1999}
{Nowak} M.~A.,  {Vaughan} B.~A.,  {Wilms} J.,  {Dove} J.~B.,   {Begelman}
  M.~C.,  1999, \mn@doi [\apj] {10.1086/306610}, \href
  {http://adsabs.harvard.edu/abs/1999ApJ...510..874N} {510, 874}

\bibitem[\protect\citeauthoryear{{Pakull} \& {Mirioni}}{{Pakull} \&
  {Mirioni}}{2003}]{pakull2003}
{Pakull} M.~W.,  {Mirioni} L.,  2003, in {Arthur} J.,  {Henney} W.~J.,  eds,
  Revista Mexicana de Astronomia y Astrofisica Conference Series Vol. 15,
  Revista Mexicana de Astronomia y Astrofisica Conference Series. pp 197--199

\bibitem[\protect\citeauthoryear{{Poutanen}, {Lipunova}, {Fabrika}, {Butkevich}
   \& {Abolmasov}}{{Poutanen} et~al.}{2007}]{poutanen2007}
{Poutanen} J.,  {Lipunova} G.,  {Fabrika} S.,  {Butkevich} A.~G.,   {Abolmasov}
  P.,  2007, \mn@doi [\mnras] {10.1111/j.1365-2966.2007.11668.x}, \href
  {http://adsabs.harvard.edu/abs/2007MNRAS.377.1187P} {377, 1187}

\bibitem[\protect\citeauthoryear{{Remillard} \& {McClintock}}{{Remillard} \&
  {McClintock}}{2006}]{remillard2006}
{Remillard} R.~A.,  {McClintock} J.~E.,  2006, \mn@doi [\araa]
  {10.1146/annurev.astro.44.051905.092532}, \href
  {http://adsabs.harvard.edu/abs/2006ARA\%26A..44...49R} {44, 49}

\bibitem[\protect\citeauthoryear{{Remillard}, {Sobczak}, {Muno}  \&
  {McClintock}}{{Remillard} et~al.}{2002}]{remillard2002}
{Remillard} R.~A.,  {Sobczak} G.~J.,  {Muno} M.~P.,   {McClintock} J.~E.,
  2002, \mn@doi [\apj] {10.1086/324276}, \href
  {http://adsabs.harvard.edu/abs/2002ApJ...564..962R} {564, 962}

\bibitem[\protect\citeauthoryear{{Scaringi}, {K{\"o}rding}, {Uttley}, {Knigge},
  {Groot}  \& {Still}}{{Scaringi} et~al.}{2012}]{scaringi2012}
{Scaringi} S.,  {K{\"o}rding} E.,  {Uttley} P.,  {Knigge} C.,  {Groot} P.~J.,
  {Still} M.,  2012, \mn@doi [\mnras] {10.1111/j.1365-2966.2012.20512.x}, \href
  {http://adsabs.harvard.edu/abs/2012MNRAS.421.2854S} {421, 2854}

\bibitem[\protect\citeauthoryear{{Shakura} \& {Sunyaev}}{{Shakura} \&
  {Sunyaev}}{1973}]{shakura1973}
{Shakura} N.~I.,  {Sunyaev} R.~A.,  1973, \aap, \href
  {http://adsabs.harvard.edu/abs/1973A\%26A....24..337S} {24, 337}

\bibitem[\protect\citeauthoryear{{Stobbart}, {Roberts}  \& {Wilms}}{{Stobbart}
  et~al.}{2006}]{stobbart2006}
{Stobbart} A.-M.,  {Roberts} T.~P.,   {Wilms} J.,  2006, \mn@doi [\mnras]
  {10.1111/j.1365-2966.2006.10112.x}, \href
  {http://adsabs.harvard.edu/abs/2006MNRAS.368..397S} {368, 397}

\bibitem[\protect\citeauthoryear{{Sutton}, {Roberts}, {Walton}, {Gladstone}  \&
  {Scott}}{{Sutton} et~al.}{2012}]{sutton2012}
{Sutton} A.~D.,  {Roberts} T.~P.,  {Walton} D.~J.,  {Gladstone} J.~C.,
  {Scott} A.~E.,  2012, \mn@doi [\mnras] {10.1111/j.1365-2966.2012.20944.x},
  \href {http://adsabs.harvard.edu/abs/2012MNRAS.423.1154S} {423, 1154}

\bibitem[\protect\citeauthoryear{{Sutton}, {Roberts}  \& {Middleton}}{{Sutton}
  et~al.}{2013}]{sutton2013}
{Sutton} A.~D.,  {Roberts} T.~P.,   {Middleton} M.~J.,  2013, \mn@doi [\mnras]
  {10.1093/mnras/stt1419}, \href
  {http://adsabs.harvard.edu/abs/2013MNRAS.435.1758S} {435, 1758}

\bibitem[\protect\citeauthoryear{{Uttley} \& {McHardy}}{{Uttley} \&
  {McHardy}}{2001}]{uttley2001}
{Uttley} P.,  {McHardy} I.~M.,  2001, \mn@doi [\mnras]
  {10.1046/j.1365-8711.2001.04496.x}, \href
  {http://adsabs.harvard.edu/abs/2001MNRAS.323L..26U} {323, L26}

\bibitem[\protect\citeauthoryear{{Uttley}}{{Uttley}}{2004}]{uttley2004}
{Uttley} P.,  2004, \mn@doi [\mnras] {10.1111/j.1365-2966.2004.07434.x}, \href
  {http://adsabs.harvard.edu/abs/2004MNRAS.347L..61U} {347, L61}

\bibitem[\protect\citeauthoryear{{Uttley}, {McHardy}  \& {Vaughan}}{{Uttley}
  et~al.}{2005}]{uttley2005}
{Uttley} P.,  {McHardy} I.~M.,   {Vaughan} S.,  2005, \mn@doi [\mnras]
  {10.1111/j.1365-2966.2005.08886.x}, \href
  {http://adsabs.harvard.edu/abs/2005MNRAS.359..345U} {359, 345}

\bibitem[\protect\citeauthoryear{{Uttley}, {Cackett}, {Fabian}, {Kara}  \&
  {Wilkins}}{{Uttley} et~al.}{2014}]{uttley2014}
{Uttley} P.,  {Cackett} E.~M.,  {Fabian} A.~C.,  {Kara} E.,   {Wilkins} D.~R.,
  2014, \mn@doi [\aapr] {10.1007/s00159-014-0072-0}, \href
  {http://adsabs.harvard.edu/abs/2014A\%26ARv..22...72U} {22, 72}

\bibitem[\protect\citeauthoryear{{Van de Sande}, {Scaringi}  \& {Knigge}}{{Van
  de Sande} et~al.}{2015}]{vandesande2015}
{Van de Sande} M.,  {Scaringi} S.,   {Knigge} C.,  2015, \mn@doi [\mnras]
  {10.1093/mnras/stv157}, \href
  {http://adsabs.harvard.edu/abs/2015MNRAS.448.2430V} {448, 2430}

\bibitem[\protect\citeauthoryear{{Vaughan} \& {Nowak}}{{Vaughan} \&
  {Nowak}}{1997}]{vaughannowak1997}
{Vaughan} B.~A.,  {Nowak} M.~A.,  1997, \mn@doi [\apjl] {10.1086/310430}, \href
  {http://adsabs.harvard.edu/abs/1997ApJ...474L..43V} {474, L43}

\bibitem[\protect\citeauthoryear{{Vaughan}, {Fabian}  \& {Nandra}}{{Vaughan}
  et~al.}{2003a}]{bvaughan2003}
{Vaughan} S.,  {Fabian} A.~C.,   {Nandra} K.,  2003a, \mn@doi [\mnras]
  {10.1046/j.1365-8711.2003.06285.x}, \href
  {http://adsabs.harvard.edu/abs/2003MNRAS.339.1237V} {339, 1237}

\bibitem[\protect\citeauthoryear{{Vaughan}, {Edelson}, {Warwick}  \&
  {Uttley}}{{Vaughan} et~al.}{2003b}]{vaughan2003}
{Vaughan} S.,  {Edelson} R.,  {Warwick} R.~S.,   {Uttley} P.,  2003b, \mn@doi
  [\mnras] {10.1046/j.1365-2966.2003.07042.x}, \href
  {http://adsabs.harvard.edu/abs/2003MNRAS.345.1271V} {345, 1271}

\bibitem[\protect\citeauthoryear{{Walton} et~al.,}{{Walton}
  et~al.}{2013}]{walton2013}
{Walton} D.~J.,  et~al., 2013, \mn@doi [\apj] {10.1088/0004-637X/779/2/148},
  \href {http://adsabs.harvard.edu/abs/2013ApJ...779..148W} {779, 148}

\bibitem[\protect\citeauthoryear{{Walton} et~al.,}{{Walton}
  et~al.}{2014}]{walton2014}
{Walton} D.~J.,  et~al., 2014, \mn@doi [\apj] {10.1088/0004-637X/793/1/21},
  \href {http://adsabs.harvard.edu/abs/2014ApJ...793...21W} {793, 21}

\bibitem[\protect\citeauthoryear{{Walton} et~al.,}{{Walton}
  et~al.}{2015a}]{walton2015b}
{Walton} D.~J.,  et~al., 2015a, \mn@doi [\apj] {10.1088/0004-637X/799/2/122},
  \href {http://adsabs.harvard.edu/abs/2015ApJ...799..122W} {799, 122}

\bibitem[\protect\citeauthoryear{{Walton} et~al.,}{{Walton}
  et~al.}{2015b}]{walton2015a}
{Walton} D.~J.,  et~al., 2015b, \mn@doi [\apj] {10.1088/0004-637X/806/1/65},
  \href {http://adsabs.harvard.edu/abs/2015ApJ...806...65W} {806, 65}

\bibitem[\protect\citeauthoryear{{Zoghbi}, {Uttley}  \& {Fabian}}{{Zoghbi}
  et~al.}{2011}]{zoghbi2011}
{Zoghbi} A.,  {Uttley} P.,   {Fabian} A.~C.,  2011, \mn@doi [\mnras]
  {10.1111/j.1365-2966.2010.17883.x}, \href
  {http://adsabs.harvard.edu/abs/2011MNRAS.412...59Z} {412, 59}

\makeatother
\end{thebibliography}

\onecolumn

\appendix

\section{rms-flux relation}

\label{rmsfluxappendix}

In this appendix we provide the plots of the rms-flux relation for NGC\,5408 X-1 and NGC\,6946 X-1 for each year separately. The results are presented in Sect. \ref{rmsflux}.

\begin{figure}
\centering
\includegraphics[width=0.35\textwidth,angle=90]{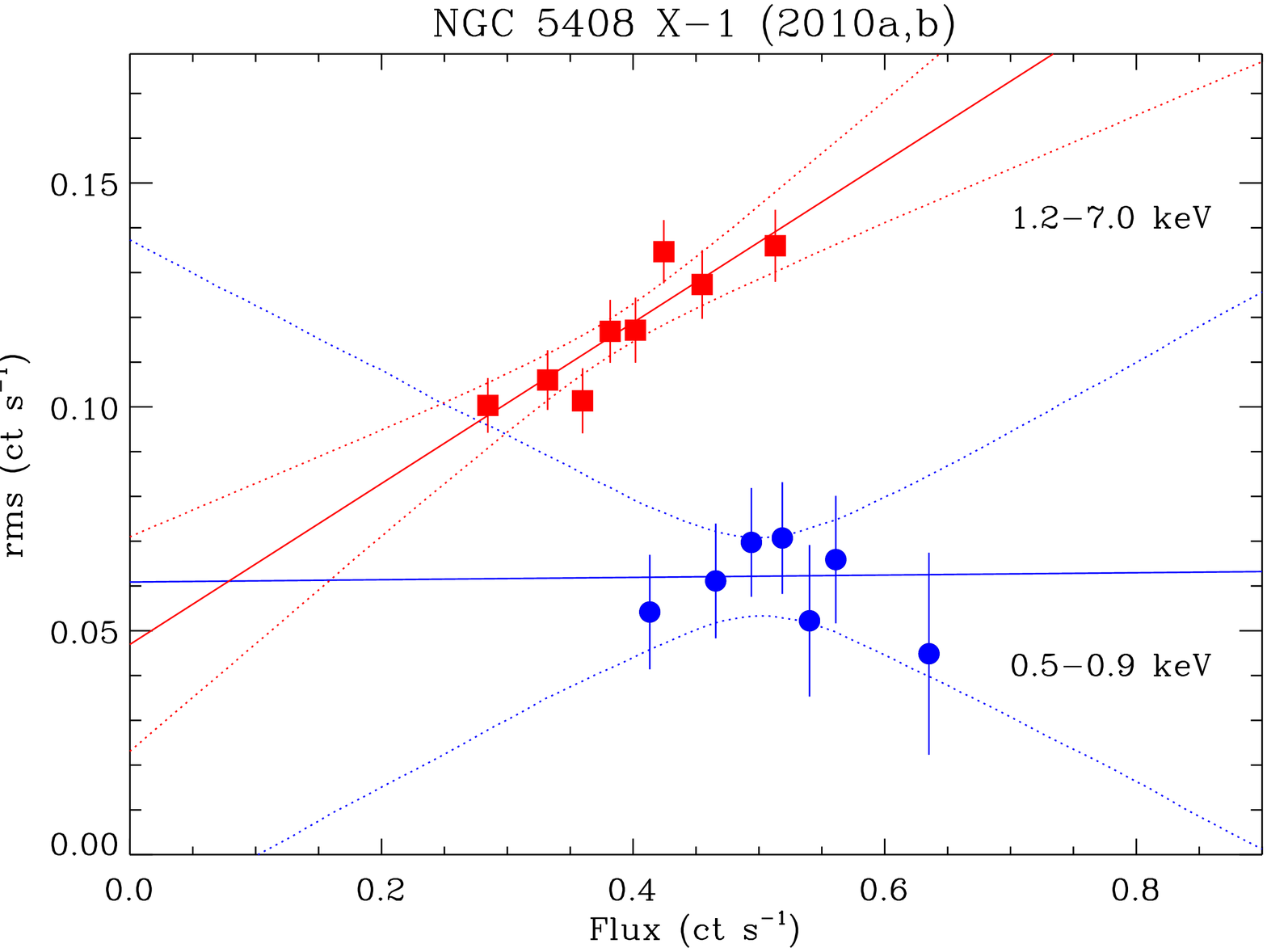}
\includegraphics[width=0.35\textwidth,angle=90]{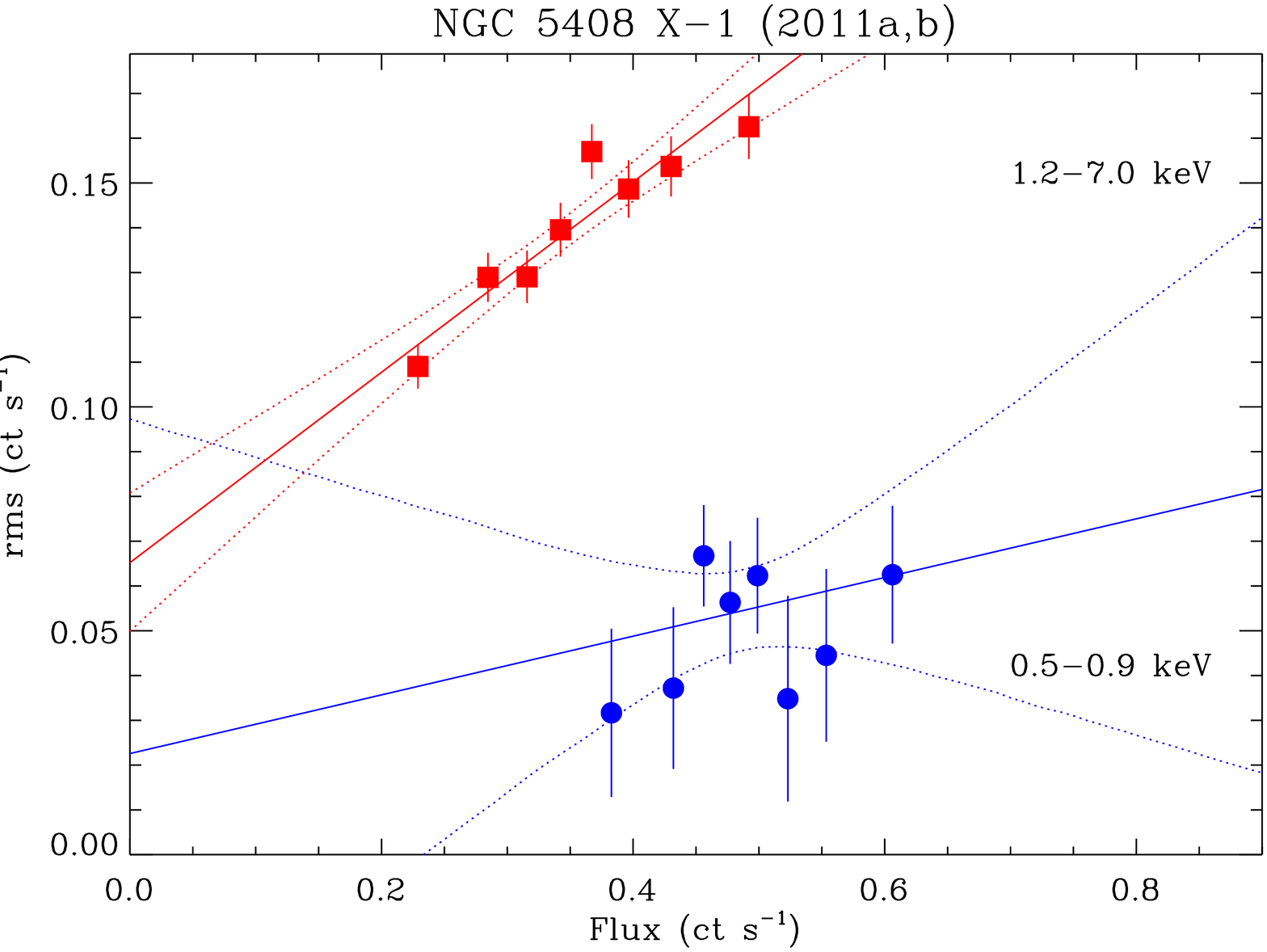}
\includegraphics[width=0.35\textwidth,angle=90]{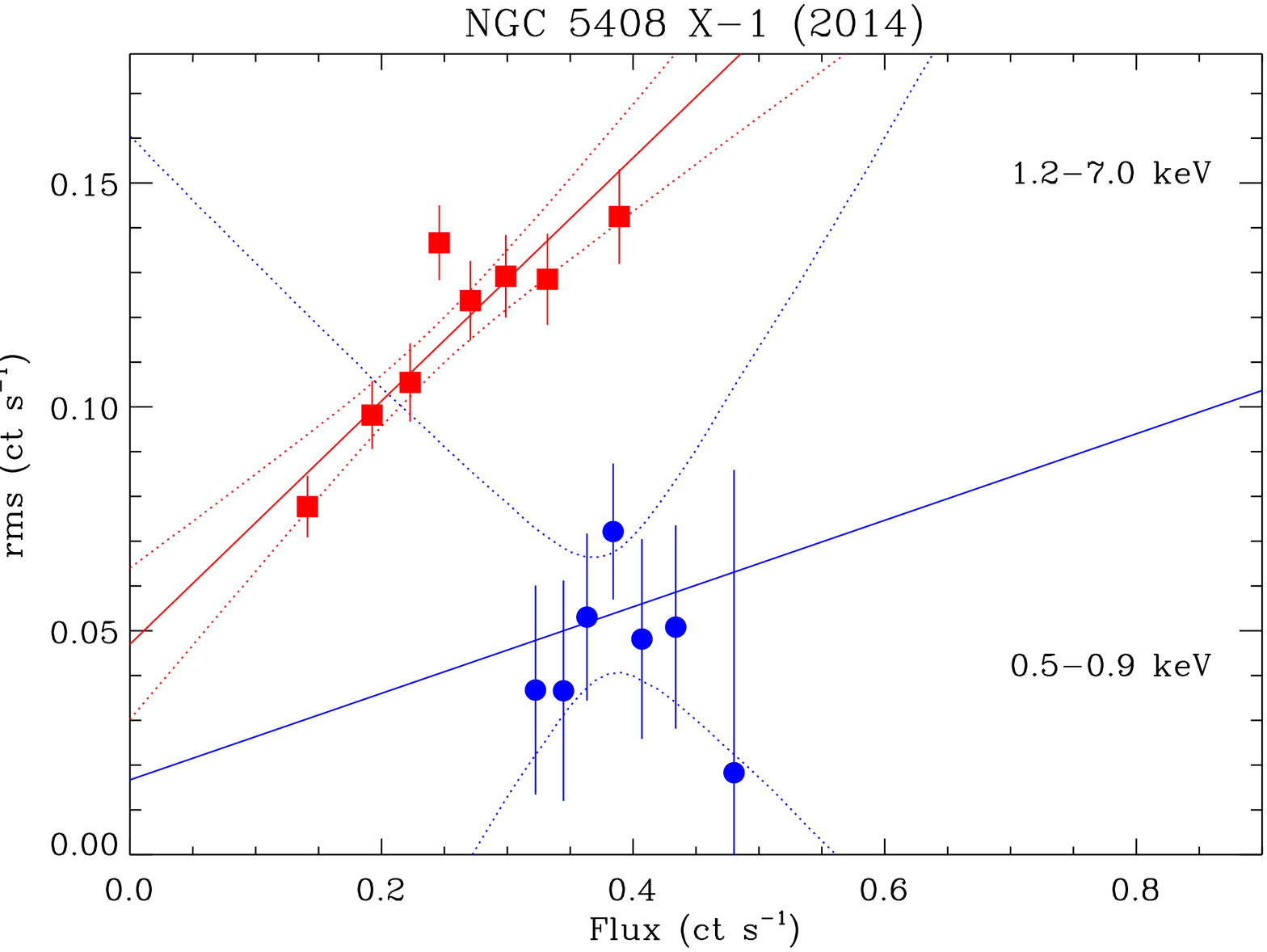}
\caption{rms-flux relation for NGC\,5408 X-1 in the soft (0.5--0.9 keV) and hard (1.2--7 keV) energy bands, for observations from 2010, 2011, and 2014. The rms is measured over the 4--50 mHz frequency range from segments of length 150 s. The dashed lines show the 95\% ``confidence bands" around the best-fitting linear model.}
\end{figure}

\begin{figure}
\centering
\includegraphics[width=0.35\textwidth,angle=90]{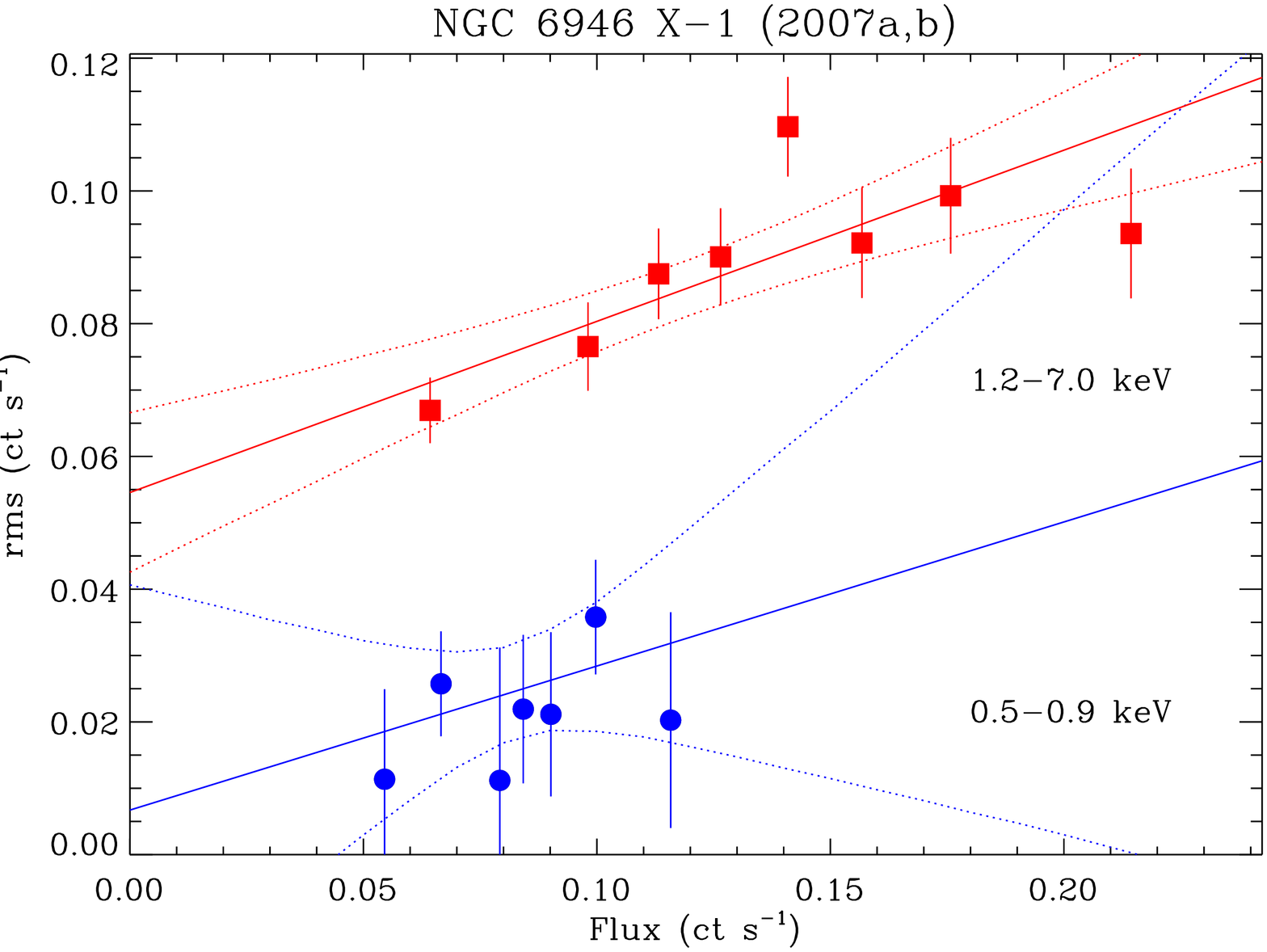}
\caption{rms-flux relation for NGC\,6946 X-1 in the soft (0.5--0.9 keV) and hard (1.2--7 keV) energy bands, for observations from 2007. The rms is measured over the 4--50 mHz frequency range from segments of length 250 s. The dashed lines show the 95\% ``confidence bands" around the best-fitting linear model.}
\end{figure}

\section{Cross-spectrum}

\label{lagsapp}

In this appendix we present the cross-spectral analysis for NGC\,6946 X-1, including the noise-subtracted coherence, phase lag, and time lag between the soft and hard energy bands. The trend of the lag is similar to that observed in NGC\,5408 X-1, although the coherence is badly constrained, and both the soft and hard lag are below the 3$\sigma$ level. Therefore, we cannot confirm the lags in this ULX.

\begin{figure}
\centering
\includegraphics[width=0.5\textwidth]{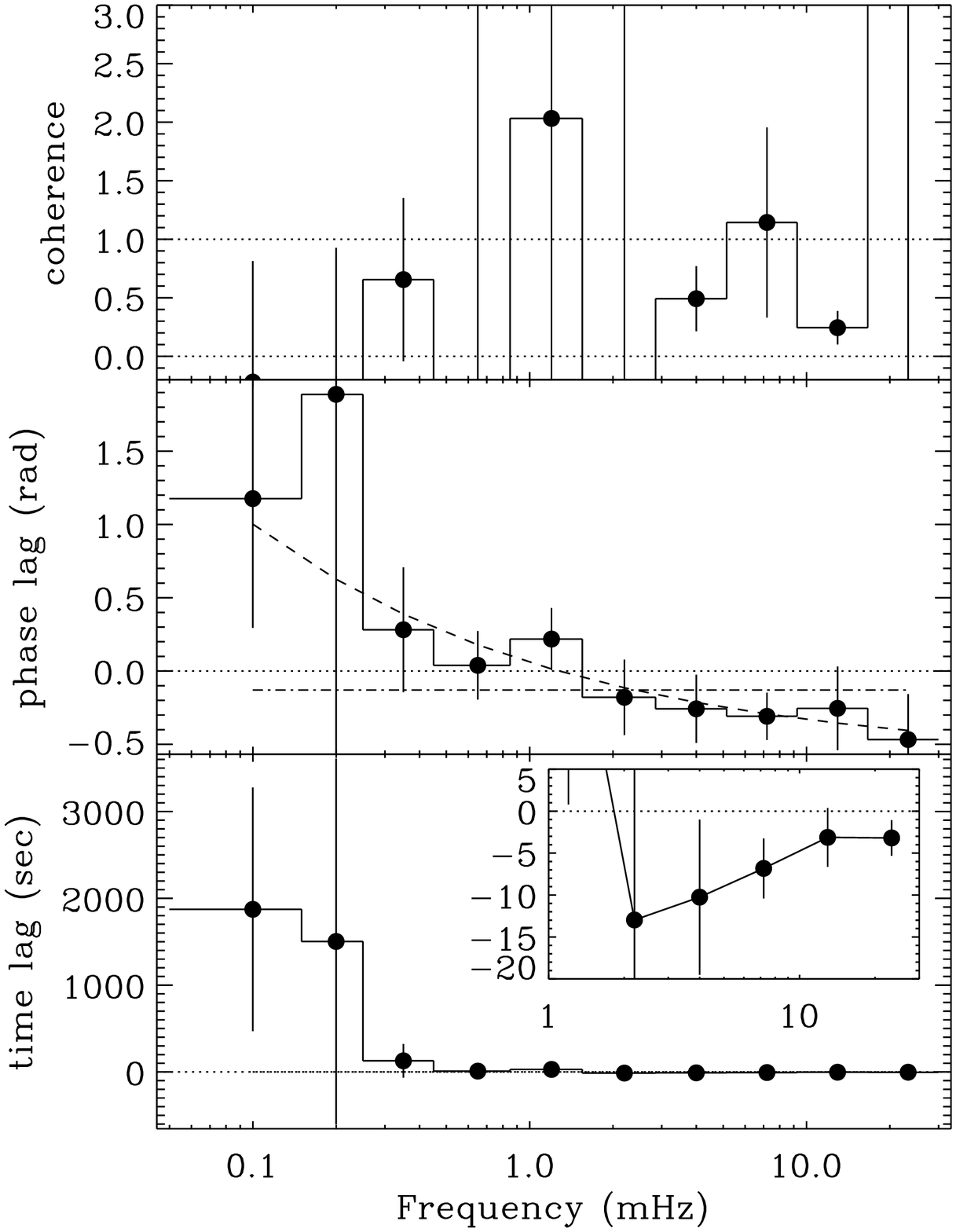}
\caption{Results of the cross-spectral analysis between the soft (0.5-0.9 keV) and hard (1.2-7.0 keV) energy bands for NGC\,6946 X-1 showing, from upper to lower panels: the noise-subtracted coherence; phase-lag; and time lag  (the inset panel shows a zoom-in of the soft lag).  The middle panel shows a constant model (dot-dashed line) and a power law plus a constant model (dashed line) we fitted to the phase lag (see text).}
\end{figure}

\end{document}